\documentclass[aps,prl,twocolumn,showpacs]{revtex4}
\usepackage{bm}
\usepackage{latexsym}
\usepackage{amssymb}
\usepackage{amsmath}
\usepackage{exscale}

\begin{document}

\title{Quantum theory of light and noise polarization in nonlinear optics}
\author{Stefan Scheel}
\email{s.scheel@imperial.ac.uk}
\affiliation{Quantum Optics and Laser Science, Blackett Laboratory,
Imperial College London, Prince Consort Road, London SW7 2BW, United Kingdom}
\author{Dirk--Gunnar Welsch}
\email{welsch@tpi.uni-jena.de}
\affiliation{Theoretisch-Physikalisches Institut,
Friedrich-Schiller-Universit\"at Jena, Max-Wien-Platz 1,
D-07743 Jena, Germany}
\date{\today}

\begin{abstract}
We present a consistent quantum theory of the electromagnetic field in
nonlinearly responding causal media, with special emphasis
on $\chi^{(2)}$ media. Starting from QED in linearly responding causal
media, we develop a method to construct the nonlinear Hamiltonian
expressed in terms of the complex nonlinear susceptibility in a
quantum mechanically consistent way. In particular we show that the
method yields the nonlinear noise polarization, which together with
the linear one is responsible for intrinsic quantum decoherence.
\end{abstract}
\maketitle
Recent advances in quantum information technologies have been
the main driving forces behind the desire to build parametric
down-conversion sources of entangled photon pairs (in the low-intensity limit)
\cite{Burnham,Kwiat,Rarity,Walmsley,Sergienko} or two-mode squeezed
states (in the high-intensity limit) \cite{Smithey}
with high fidelity. It is known that single-photon states of non-unit
efficiency as produced by heralded single-photon sources using
parametric down-conversion, cannot be purified by using linear optical
elements and photo detection to yield states with higher efficiency
\cite{Berry1,Berry2}. That in turn means that post-processing of
single-photon sources is impossible and the sources themselves have to
be improved. In order to achieve the maximally possible purity of
heralded single-photon states or correlated (entangled) twin-beam
photons it is therefore necessary to investigate the theoretical
limits nature imposes on us.

An important step in this direction is to provide a quantum theory of
light that takes into account nonlinear processes
such as parametric down-conversion,
and at the same time decoherence mechanisms due to
unavoidable absorption losses of the nonlinear material the
light interacts with. The theory of
quantized electromagnetic fields in linearly and causally responding
materials (with the linear response function satisfying the
Kramers--Kronig relations) is well established
(see, e.g., Refs.~\cite{HuttnerBarnett,Buch,Suttorp}). It has been known
for some time that analogous Kramers--Kronig relations do also
hold for nonlinear susceptibilities \cite{Bloembergen}. Hence, it
will be interesting to see how these causal relations
appear in a nonlinear quantum theory.

Previous work on electromagnetic field quantization in nonlinear
materials have focused on strictly lossless materials where
Lagrangian methods and mode decompositions apply
\cite{Hillery84,Drummond90,Abram91,Duan97}.
A first attempt to include in the field quantization
both linear and nonlinear losses was made
in Ref.~\cite{Schmidt98} for Kerr media, by extending the
linear harmonic-oscillator model used in the
Huttner--Barnett quantization scheme \cite{HuttnerBarnett}
to a nonlinear one.
A consistent approach that includes---for given nonlinear
susceptibility---absorption and dispersion has not yet been formulated
within the frame of (macroscopic) QED.

In this article we will exemplify, on focusing on $\chi^{(2)}$ media,
how to consistently quantize the electromagnetic field in the presence of
nonlinearly responding causal materials.
This theory provides the starting point
for further investigations of theoretical limits to the performance of
nonlinear optical elements as sources of nonclassical light.
Starting from the nonlinear Hamiltonian expressed in terms of the
canonically conjugated variables as used in QED in linear causal media,
we first express the nonlinear polarization
field in terms of these variables as well. This is compared with the
classical nonlinear response which enables us to identify the
nonlinear noise contributions.

We begin with recapitulating the quantization scheme for the
electromagnetic field in the presence of a linearly (and locally)
responding causal dielectric medium
of permittivity $\varepsilon(\mathbf{r},\omega)$
$\!=$ $\!\varepsilon'(\mathbf{r},\omega)$
$\!+$ $\!i\varepsilon''(\mathbf{r},\omega)$
\cite{Buch}. In this case the Hamiltonian is bilinear,
\begin{equation}
\label{eq:hamlin}
H_L = \int d^3r \int\limits_0^\infty d\omega \,\hbar \omega\,
\mathbf{f}^\dagger(\mathbf{r},\omega) \cdot \mathbf{f}(\mathbf{r},\omega),
\end{equation}
with the annihilation and creation operators
$f_i(\mathbf{r},\omega)$ and $f_i^\dagger(\mathbf{r},\omega)$, respectively,
playing the role of the canonically conjugated dynamical variables
which are attributed to collective excitations of the electromagnetic
field and the dispersing and absorbing dielectric matter and
obeying the bosonic commutation rules
$[f_i(\mathbf{r},\omega), f_j^\dagger(\mathbf{r}',\omega')]=
\linebreak
\delta_{ij} \delta(\omega-\omega')
\delta(\mathbf{r}-\mathbf{r}')$.
By expressing the electromagnetic field in terms of the dynamical
variables, the electric field, for example, reads
\begin{gather}
\label{eq:efeld-0}
\mathbf{E}(\mathbf{r}) = \int_0^\infty d\omega\,
\underline{\mathbf{E}}(\mathbf{r},\omega)
+ \mathrm{h.c.},
\\
\label{eq:efeld}
\underline{\mathbf{E}}(\mathbf{r},\omega)
= i\sqrt{\frac{\hbar}{\pi\varepsilon_0}}
\frac{\omega^2}{c^2} \!\int\! d^3s
\sqrt{\varepsilon''(\mathbf{s},\omega)}\,
\bm{G}(\mathbf{r},\mathbf{s},\omega)
\cdot\mathbf{f}(\mathbf{s},\omega) ,
\end{gather}
it can then be shown that Maxwell's equations,
in particular Faraday's and Ampere's laws, hold.
In Eq.~(\ref{eq:efeld}) the (dyadic) Green function
$\bm{G}(\mathbf{r},\mathbf{s},\omega)$ is the unique
fundamental solution of the inhomogeneous Helmholtz equation
\begin{equation}
\label{eq:helmholtz}
\bm{\nabla}\times\bm{\nabla}\times
\bm{G}(\mathbf{r},\mathbf{s},\omega) -\frac{\omega^2}{c^2}
\varepsilon(\mathbf{r},\omega) \bm{G}(\mathbf{r},\mathbf{s},\omega)
= \delta(\mathbf{r}-\mathbf{s}) \bm{I}
\end{equation}
and contains all relevant information about the material properties
and the geometry of the system.

Equation (\ref{eq:efeld-0}) together with Eq.~(\ref{eq:efeld})
may be regarded as a generalization
of the ordinary mode expansion, with the role of the mode operators
being taken on, in a sense, by the $f_i(\mathbf{r},\omega,t)$ and
$f_i^\dagger(\mathbf{r},\omega,t)$.
In summary, (i) the Hamiltonian (\ref{eq:hamlin}) generates, using the
representation according to Eqs.~(\ref{eq:efeld-0}) and (\ref{eq:efeld}),
the correct (macroscopic) Maxwell equations, (ii) the fundamental QED
equal-time commutation relations are preserved,
and (iii) the fluctuation-dissipation theorem takes its standard form.
Moreover, the Hamiltonian (\ref{eq:hamlin}) represents an energy
stored in the system composed of the electromagnetic field and
absorbing matter.

To turn over to the nonlinear media, let us first fix some
notation. From now on we will abbreviate spatial
and frequency variables $(\mathbf{r}_k,\omega_k)$ by their label
$\bm{k}$, e.g., \mbox{$\bm{1}$ $\!\equiv$ $\!(\mathbf{r}_1,\omega_1)$}
and write
$\int d\bm{k}\equiv$ $\!\int d^3r_k\int d\omega_k$. In the latter
integrals, the spatial integration extends over all space. The
frequency integral, which we initially will assume to range over all
positive frequencies, will be restricted later on.
On recalling the physical meaning of the dynamical variables
$f_i(\mathbf{r},\omega)$ and $f_i^\dagger(\mathbf{r},\omega)$,
the most general normal-order form of the nonlinear interaction
energy that corresponds to a $\chi^{(2)}$ medium reads
\begin{equation}
\label{eq:hamnonlin}
H_{NL} = \!\int\! d\bm{1}\,d\bm{2}\,d\bm{3}\,
\alpha_{i(jk)}(\bm{1},\bm{2},\bm{3})
f_i^\dagger(\bm{1}) f_j(\bm{2}) f_k(\bm{3})
+\mbox{h.c.}.
\end{equation}
The unknown tensor function
$\alpha_{i(jk)}(\bm{1},\bm{2},\bm{3})$, which has to be symmetrized
over its last two indices to avoid double-counting, has to be
determined from constraints imposed by generally accepted relations.

We first note that Faraday's law can be written as
\begin{equation}
\bm{\nabla}\times \mathbf{E}(\mathbf{r}) =
-\dot{\mathbf{B}}(\mathbf{r}) = -\frac{1}{i\hbar} \left[
\mathbf{B}(\mathbf{r}), H_L+H_{NL} \right].
\end{equation}
Both the electric and the magnetic induction fields are pure
electromagnetic fields without being related to the material degrees
of freedom and
hence their equal-time commutation relations are as in vacuum QED.
To be more specific, we may assume that the functional form of these
fields in terms of the dynamical variables
$f_(\mathbf{r},\omega)$ and $f_i^\dagger(\mathbf{r},\omega)$
is (in close analogy to the case of ordinary mode expansion)
the same as in the linear theory. From here it immediately
follows that
\begin{equation}
\label{eq:constraint1}
\left[ \mathbf{B}(\mathbf{r}), H_{NL} \right] = 0 .
\end{equation}

Using Faraday's law, we rewrite Ampere's law as
\begin{equation}
\label{eq:Faraday}
\bm{\nabla}\times\bm{\nabla}\times\mathbf{E}(\mathbf{r}) = -\mu_0
\ddot{\mathbf{D}}_L(\mathbf{r}) -\mu_0
\ddot{\mathbf{P}}_{NL}
(\mathbf{r}),
\end{equation}
where we have split up the dielectric displacement field
$\mathbf{D}(\mathbf{r})$ into the linear part
$\mathbf{D}_L(\mathbf{r})$ and the nonlinear polarization
$\mathbf{P}_{NL}(\mathbf{r})$. Employing Heisenberg's equation of
motion, we may rewrite Eq.~(\ref{eq:Faraday}) as
\begin{align}
\label{eq:constraint2}
&\bm{\nabla}\times\bm{\nabla}\times\mathbf{E}(\mathbf{r})
-\frac{\mu_0}{\hbar^2} \bigl[ \left[ \mathbf{D}_L(\mathbf{r}),H_L
\right] ,H_L \bigr]
\nonumber \\ &
=\frac{\mu_0}{\hbar^2} \left\{
\bigl[ \left[ \mathbf{D}_L(\mathbf{r}),H_L \right] ,H_{NL} \bigr]
+\bigl[ \left[ \mathbf{D}_L(\mathbf{r}),H_{NL} \right] ,H_L \bigr]
\right.
\nonumber \\ &
\left.
+\bigl[ \left[
\mathbf{P}_{NL}
(\mathbf{r}),H_L \right] ,H_L \bigr]
\right\},
\end{align}
where we have kept,
for consistency reasons,
only terms that are in at most first order in the
nonlinear coupling coefficient $\alpha_{i(jk)}(\bm{1},\bm{2},\bm{3})$.
The lhs of Eq.~(\ref{eq:constraint2}) is zero by the definition of the
linear displacement field. Note that the time dependence is carried by
the time-dependent dynamical variables $f_i(\mathbf{r},\omega,t)$ and
$f_i^\dagger(\mathbf{r},\omega,t)$.
The first term on the rhs of Eq.~(\ref{eq:constraint2}) vanishes by
virtue of the constraint (\ref{eq:constraint1}). To see this, one has to
express the linear displacement and the magnetic induction fields
in terms of the dynamical variables,
leading to $[\mathbf{D}_L(\mathbf{r}),H_L]$ $\!=$
$\!(i\hbar/\mu_0)\bm{\nabla}\times\mathbf{B}(\mathbf{r})$,
and application of Eq.~(\ref{eq:constraint1}) leads to the quoted
result.
Hence, we are left with a relation between double commutators of the
linear displacement and nonlinear polarization fields
with the linear and nonlinear parts of the Hamiltonian,
$[[\mathbf{D}_L(\mathbf{r}),H_{NL}],H_L]$ $\!=$
$\!-[[\mathbf{P}_{NL}(\mathbf{r}),H_L],H_L]$.
A particular solution is certainly
\begin{equation}
\label{eq:constraint4}
\left[ \mathbf{D}_L(\mathbf{r}),H_{NL} \right] =
- \left[ \mathbf{P}_{NL}(\mathbf{r}),H_L \right] .
\end{equation}
The general solution would additionally include commutants with the
linear Hamiltonian $H_L$. These terms must be functionals of the
number (density) operator
$\mathbf{f}^\dagger(\mathbf{r},\omega) \cdot \mathbf{f}(\mathbf{r},\omega)$.
However, we can assume that linear functionals of this type are
already included in the particular solution (\ref{eq:constraint4}) as
they lead to bilinear forms in the dynamical variables. On the other
hand, quartic and higher functionals have to be excluded to ensure
that $\mathbf{P}_{NL}(\mathbf{r})$ stays bilinear which guarantees
consistency within the approximations made.

The expression on the rhs of Eq.~(\ref{eq:constraint4}) is
nothing but the Liouvillian ${\cal L}_L$ generated by the linear
Hamiltonian $H_L$ acting on the nonlinear polarization
field. Therefore, Eq.~(\ref{eq:constraint4}) can be
solved for $\mathbf{P}_{NL}(\mathbf{r})$ to yield
\begin{equation}
\label{eq:10}
\mathbf{P}_{NL}
(\mathbf{r}) = -\frac{1}{i\hbar} {\cal L}_L^{-1}
\left[ \mathbf{D}_L(\mathbf{r}),H_{NL} \right] .
\end{equation}
At this point we recall that according to
\begin{eqnarray}
\label{eq:dl}
\underline{\mathbf{D}}_L(\mathbf{r},\omega) &=&
(\mu_0\omega^2)^{-1}\bm{\nabla}\times\bm{\nabla}\times
\underline{\mathbf{E}}(\mathbf{r},\omega) \nonumber \\
&=& \varepsilon_0 \varepsilon(\mathbf{r},\omega)
\underline{\mathbf{E}}(\mathbf{r},\omega)
+\underline{\mathbf{P}}_L^{(N)}(\mathbf{r},\omega)
\end{eqnarray}
the linear displacement field
\begin{equation}
\label{eq:dl-1}
\mathbf{D}_L(\mathbf{r})= \int_0^\infty d\omega\,
\underline{\mathbf{D}}_L(\mathbf{r},\omega)+\mbox{h.c.}
\end{equation}
consists of a reactive part related to the electric field
and a noise part
\begin{equation}
\label{eq:dl-2}
\mathbf{P}_L^{(N)}(\mathbf{r})= \int_0^\infty d\omega\,
\underline{\mathbf{P}}_L^{(N)}(\mathbf{r},\omega)+\mbox{h.c.}.
\end{equation}
Inserting Eq.~(\ref{eq:dl-1}) together with Eq.~(\ref{eq:dl}) into
Eq.~(\ref{eq:10}), we see that the nonlinear polarization also
decomposes into a reactive part, which can be related to the nonlinear
response, and a noise part, which determines the nonlinear noise
polarization
\begin{equation}
\mathbf{P}_{NL}^{(N)}
(\mathbf{r}) = -\frac{1}{i\hbar} {\cal L}_L^{-1}
\left[ \mathbf{P}_L^{(N)}(\mathbf{r}),H_{NL} \right] .
\end{equation}
Because of the relation $\underline{\mathbf{P}}_L^{(N)}(\mathbf{r},\omega)$
$\!=i\sqrt{\hbar\varepsilon_0/\pi}\sqrt{\varepsilon''(\textbf{r},\omega)}$
$\textbf{f}(\textbf{r},\omega)$, $\mathbf{P}_{NL}^{(N)}(\mathbf{r})$
vanishes if the imaginary part of the linear permittivity,
$\varepsilon''(\textbf{r},\omega)$, and hence the noise associated
with it tends to zero \footnote{This argument does not apply to
Eq.~(\ref{eq:efeld}) as changing the order of integration
and taking the limit is not allowed.}.

The inverse Liouvillian can be calculated using standard
techniques, and we obtain from Eq.~(\ref{eq:10})
\begin{align}
\label{eq:dnl}
&
\mathbf{P}_{NL}
(\mathbf{r}) =
\nonumber\\&\quad
\frac{i}{\hbar} \lim_{s\to 0}
\int\limits_0^\infty \!d\tau \,e^{-s\tau}
e^{-\frac{i}{\hbar}H_L\tau}
\left[ \mathbf{D}_L(\mathbf{r}),H_{NL} \right]
e^{\frac{i}{\hbar}H_L\tau},
\end{align}
where the real positive number $s$ ensures convergence of the integral.
In the next step we compute the commutator
$[\mathbf{D}_L(\mathbf{r}),H_{NL}]$ and evaluate the integral in
Eq.~(\ref{eq:dnl}),
First, we evaluate the
commutator between the dynamical variables and the nonlinear Hamiltonian
$H_{NL}$, leading to [here, $\bm{0}\equiv(\mathbf{s},\omega)$]
\begin{align}
\label{eq:comm1}
&\left[ f_m(\bm{0}),H_{NL} \right]
=\int d\bm{2}d\bm{3}\,\alpha_{m(jk)}(\bm{0},\bm{2},\bm{3})
f_j(\bm{2}) f_k(\bm{3})
\nonumber \\&\qquad
+\int d\bm{1}d\bm{2}\,\alpha_{i(jm)}^\ast(\bm{1},\bm{2},\bm{0})
f_j^\dagger(\bm{2}) f_i(\bm{1}) .
\end{align}
In what follows, we will concentrate on the contribution to the
nonlinear displacement and polarization that comes from terms
containing two annihilation operators such as
$f_j(\bm{2})f_k(\bm{3})$. We will label these contributions with the
superscript ${}^{(++)}$ in analogy with the standard notation for
positive-frequency parts.
The inverse Liouvillian of the bilinear combination of
annihilation operators is readily found to be
${\cal L}^{-1}f_j(\bm{2})f_k(\bm{3})=$
$i/(\omega_2+\omega_3)f_j(\bm{2})f_k(\bm{3})$. Combined with
Eq.~(\ref{eq:dnl}) we finally obtain for the nonlinear
polarization field
\begin{align}
\label{eq:dnl2}
&
P_{NL,l}^{(++)}
(\mathbf{r}) =
\frac{1}{i\hbar} \sqrt{\frac{\hbar\varepsilon_0}{\pi}}
\!\int\! d\bm{0}d\bm{2}d\bm{3}\,
\frac{\sqrt{\varepsilon''(\bm{0})}}{\omega_2+\omega_3}
\alpha_{m(jk)}(\bm{0},\bm{2},\bm{3})
\nonumber \\ &\
\times\,
\frac{\omega^2}{c^2}\varepsilon(\mathbf{r},\omega)\,
G_{lm}(\mathbf{r},\bm{0})
f_j(\bm{2}) f_k(\bm{3})
+P_{NL,l}^{(N,++)}(\mathbf{r}),
\end{align}
where the noise polarization reads
\begin{align}
\label{eq:dnl2-1}
&
P_{NL,l}^{(N,++)}(\mathbf{r}) =
\frac{1}{i\hbar} \sqrt{\frac{\hbar\varepsilon_0}{\pi}}
\!\int\! d\bm{0}d\bm{2}d\bm{3}\,
\frac{\sqrt{\varepsilon''(\bm{0})}}{\omega_2+\omega_3}
\alpha_{l(jk)}(\bm{0},\bm{2},\bm{3})
\nonumber \\ &\hspace{25ex}
\times\,
\delta(\textbf{r}-\textbf{s})
f_j(\bm{2}) f_k(\bm{3}) .
\end{align}

In order to make contact with standard notation,
let us recall the definition of the nonlinear
polarization within the framework of response theory:
\begin{align}
\label{eq:zeitraum}
&P_{NL,l}(\mathbf{r},t) =
\varepsilon_0 \int\limits_{-\infty}^t d\tau_1
d\tau_2 \,
\check{\chi}
^{(2)}_{lmn}(\mathbf{r},t-\tau_1,t-\tau_2)
\nonumber \\ &\hspace{15ex}
\times
E_m(\mathbf{r},\tau_1) E_n(\mathbf{r},\tau_2)
+ P_{NL,l}^{(N)}(\mathbf{r},t)
.
\end{align}
The first term on the rhs of Eq.~(\ref{eq:zeitraum}) is the
causal response well known from nonlinear optics
\cite{SchubertWilhelmi}, with $\check{\chi}^{(2)}_{lmn}(\mathbf{r},t_1,t_2)$
being the response function of the $\chi^{(2)}$ medium.
The term $P_{NL,l}^{(N)}(\mathbf{r},t)$ is a (yet unknown) nonlinear
noise polarization commonly disregarded in classical nonlinear optics.
In most cases of interest it is sufficient to evaluate
Eq.~(\ref{eq:zeitraum}) in the slowly-varying amplitude approximation
in the sense that
\begin{align}
\label{eq:15}
&\mathbf{E}(\mathbf{r},t) = \sum_{\nu=1}^3
\tilde{\mathbf{E}}_i(\mathbf{r},\Omega_\nu,t)e^{-i\Omega_\nu t}
+ \mathrm{h.c.},
\end{align}
with the time scale on which the amplitude function
$\tilde{E}_i(\mathbf{r},\Omega_\nu,t)$ noticeably changes being
long compared with $\Omega_\nu^{-1}$ and the characteristic time
of variation
of $\chi^{(2)}_{lmn}(\mathbf{r},t_1,t_2)$ with respect to
both $t_1$ and $t_2$ (see, e.g., the treatment in Ref.~\cite{Kovsh}).
Hence the slowly varying field amplitudes can be taken
out of the integral at the upper integration limit $t$,
and we are left with the Fourier transform
of  $\check{\chi}^{(2)}_{lmn}(\mathbf{r},t_1,t_2)$,
$\chi^{(2)}_{lmn}(\mathbf{r},\omega_1,\omega_2)$,
which slowly varies with $\omega_1$ and $\omega_2$. In this way we derive
\begin{align}
\label{eq:pnl}
&\tilde{P}_{NL,l}^{(++)}(\mathbf{r},\Omega_{23}) = \varepsilon_0
\chi^{(2)}_{lmn}(\mathbf{r},\Omega_2,\Omega_3)
\nonumber \\ & \hspace{5ex}\times\,
\tilde{E}_m(\mathbf{r},\Omega_2) \tilde{E}_n(\mathbf{r},\Omega_3)
+
\tilde{P}_{NL,l}^{(N)}(\mathbf{r},\Omega_{23})
\end{align}
[$\Omega_{23}$ $\!\equiv$ $\!\Omega_2$ $\!+$ $\!\Omega_3$],
where the time argument $t$ of the $\tilde{}$ quantities
has been omitted for notational convenience.

The validity of the approximation leading from
Eq.~(\ref{eq:zeitraum}) to Eq.~(\ref{eq:pnl}) may be regarded
as being a prerequisite for substantiating the effective
interaction Hamiltonian (\ref{eq:hamnonlin}). At the same time, it
suggests further specification of the Hamiltonian as therein the
introduction of slowly varying variables is desirable. In view
of Eqs.~(\ref{eq:efeld-0}) and (\ref{eq:efeld}), we define,
on assuming the Green tensor and the linear susceptibility
are slowly varying with $\omega$, the slowly varying bosonic variables
$\tilde{\mathbf{f}}(\mathbf{r},\Omega_\nu)$ $\!=$
$\!(\Delta\Omega_\nu)^{-1/2}$
$\int_{\Delta\Omega_\nu} d\omega\,\mathbf{f}(\mathbf{r},
\omega,t)e^{i\Omega_\nu t}$ ($\Delta\Omega_\nu$, relevant frequency
interval around $\Omega_\nu$), and Eq.~(\ref{eq:hamnonlin}) reduces to
\begin{align}
\label{eq:hamnonlin-1}
& H_{NL} = \int d^3s_1 d^3s_2 d^3s_3\,
\alpha_{i(jk)}(\mathbf{s}_1,\Omega_{23},\mathbf{s}_2,\Omega_2,\mathbf{s}_3,\Omega_3)
\nonumber\\
& \sqrt{\Delta\Omega_1\Delta\Omega_2\Delta\Omega_3}\,
\tilde{f}_i^\dagger(\mathbf{s}_1,\Omega_{23})
\tilde{f}_j(\mathbf{s}_2,\Omega_2)\tilde{f}_k(\mathbf{s}_3,\Omega_3)
+\mathrm{h.c.}.
\end{align}

Introducing in Eqs.~(\ref{eq:dnl2}) and (\ref{eq:pnl}) the slowly varying
variables $\tilde{\mathbf{f}}(\mathbf{r},\Omega_\nu)$, from a comparison
of the reactive parts of the nonlinear polarization as given by the
two equations we derive the following integral equation for
determining the nonlinear coupling coefficient
$\alpha_{i(jk)}(\mathbf{s}_1,\Omega_{23},
\mathbf{s}_2,\Omega_2;\mathbf{s}_3,\Omega_3)$ in terms of the
nonlinear susceptibility
$\chi^{(2)}_{lmn}(\mathbf{r},\Omega_2,\Omega_3)$:
\begin{align}
\label{eq:alpha}
&\int d^3s\, \sqrt{\varepsilon''(\mathbf{s},\Omega_{23})}\,
\alpha_{m(jk)}(\mathbf{s},\Omega_{23},
\mathbf{s}_2,\Omega_2;\mathbf{s}_3,\Omega_3)
\nonumber\\&\hspace{2.5ex}\times\,
G_{lm}(\mathbf{r},\mathbf{s},\Omega_{23})
=\frac{\hbar^2}{i\pi c^2}\sqrt{\frac{\pi}{\hbar\varepsilon_0}}\,
\frac{\Omega_2^2\Omega_3^2}{\Omega_{23}}
\nonumber\\&
\hspace{2.5ex}\times\,
\sqrt{\varepsilon''(\mathbf{s}_2,\Omega_2)
\varepsilon''(\mathbf{s}_3,\Omega_3)}\,
\chi^{(2)}_{lmn}(\mathbf{r},\Omega_2,\Omega_3)
\nonumber \\ & \hspace{2.5ex}\times\,
G_{mj}(\mathbf{r},\mathbf{s}_2,\Omega_2)
G_{nk}(\mathbf{r},\mathbf{s}_3,\Omega_3) .
\end{align}
This equation is of Fredholm type and can be solved by inverting the
integral kernel on the lhs of Eq.~(\ref{eq:alpha}).
Note that the inverse of the Green tensor is just the Helmholtz
operator $H_{ij}(\mathbf{r},\omega)$
$\!=$ $\!\partial^r_i\partial^r_j$ $\!-$ $\!\delta_{ij}\Delta^r$
$\!-$ $\!(\omega^2/c^2)\varepsilon(\mathbf{r},\omega)\delta_{ij}$:
$H_{ij}(\mathbf{r},\omega) G_{jk}(\mathbf{r},\mathbf{s},\omega)$
$\!=$ $\!\delta_{ik}\delta(\mathbf{r}$ $\!-$ $\!\mathbf{s})$.
Hence, from Eq.~(\ref{eq:alpha}) it follows that
\begin{align}
\label{eq:alpha2}
&\alpha_{i(jk)}(\mathbf{r},\Omega_{23},
\mathbf{s}_2,\Omega_2;\mathbf{s}_3,\Omega_3) =
\frac{\hbar^2}{i\pi c^2}\sqrt{\frac{\pi}{\hbar\varepsilon_0}}\,
\frac{\Omega_2^2\Omega_3^2}{\Omega_{23}}
\nonumber \\ & \hspace{1ex} \times\,
\sqrt{\frac{\varepsilon''(\mathbf{s}_2,\Omega_2)
\varepsilon''(\mathbf{s}_3,\Omega_3)}
{\varepsilon''(\mathbf{r},\Omega_{23})}}
\frac{1}{\varepsilon(\mathbf{r},\Omega_{23})}\,
H_{li}(\mathbf{r},\Omega_{23})
\nonumber \\ & \hspace{1ex} \times\,
\bigl[
\chi^{(2)}_{imn}(\mathbf{r},\Omega_2,\Omega_3)
G_{mj}(\mathbf{r},\mathbf{s}_2,\Omega_2)
G_{nk}(\mathbf{r},\mathbf{s}_3,\Omega_3) \bigr] .
\end{align}

Re-inserting Eq.~(\ref{eq:alpha2}) into Eq.~(\ref{eq:dnl2-1})
eventually yields, on recalling Eqs.~(\ref{eq:efeld-0}),
(\ref{eq:efeld}), (\ref{eq:15}) and the definition of the slowly
varying variables $\tilde{\mathbf{f}} (\mathbf{r},\Omega_\nu)$,
the following expression for the nonlinear noise polarization:
\begin{align}
\label{eq:pnlnoise}
&\tilde{P}_{NL,l}^{(N,++)}(\mathbf{r},\Omega_{23}) =
\frac{\varepsilon_0c^2}{\Omega_{23}^2\varepsilon(\mathbf{r},\Omega_{23})}
\nonumber \\ &\ \hspace*{-3ex} \times 
H_{li}(\mathbf{r},\Omega_{23})
\bigl[
\chi^{(2)}_{imn}(\mathbf{r},\Omega_2,\Omega_3)
\tilde{E}_m(\mathbf{r},\Omega_2) \tilde{E}_n(\mathbf{r},\Omega_3) \bigr] .
\end{align}
To our knowledge, this is the first time a nonlinear noise
polarization has been derived in the frame of
quantum nonlinear optics. Note that the Helmholtz operator acting on
the electric field returns the linear noise polarization, 
$H_{ij}(\mathbf{r},\omega)\underline{E}_j(\mathbf{r},\omega)$ $\!=$
$\!\omega^2/(\varepsilon_0c^2)\underline{P}_{L,i}^{(N)}(\mathbf{r},\omega)$
[cf. Eq.~(\ref{eq:dl})]. Among other terms, Eq.~(\ref{eq:pnlnoise})
contains products of the electric field and the linear noise polarization.

In summary, we have presented a consistent quantum theory of the
electromagnetic field in the presence of quadratically responding
dielectric materials. It takes care of the causal nature of the
dielectric response which implies the existence of a nonlinear noise
polarization. The nonlinear (effective) interaction Hamiltonian
(\ref{eq:hamnonlin-1})
[or equivalently, Eq.~(\ref{eq:hamnonlin}) in the slowly-varying
amplitude approximation],
together with the nonlinear coupling coefficient
from Eq.~(\ref{eq:alpha2}) allows one to study nonlinear quantum
optical processes such as parametric down-conversion in the presence
of realistic dielectric materials. The main advantage of our approach
is that it automatically takes absorption---via the complex
permittivity---and geometric boundaries---via the dyadic Green
function---into account. The procedure to generalize the theory
presented above is by no means restricted to quadratic responses. In
fact, one can construct a hierarchy of Hamiltonians with increasing
number of the dynamical variables $\mathbf{f}(\mathbf{r},\omega)$ and
$\mathbf{f}^\dagger(\mathbf{r},\omega)$ corresponding to higher-order
nonlinear responses. The construction ensures that the
equal-time commutation relations between the relevant field operators
are preserved. We believe this theory represents an important step
towards further studies with the aim to understand the ultimate
limits on the performance of quantum optical processes.

\acknowledgments
This work was funded by the UK Engineering and Physical
Sciences Research Council (EPSRC). The authors thank A.~Tip for
helpful discussions.


\end{document}